\documentclass[aps, prd, letterpaper, 12pt, nofootinbib, superscriptaddress, longbibliography, notitlepage]{revtex4-1}
\usepackage[utf8]{inputenc}
\usepackage{amsmath,amssymb,amsfonts}
\usepackage{mathrsfs}
\usepackage{color}
\usepackage{graphicx} 
\usepackage[section]{placeins}
\usepackage[colorlinks=true,linkcolor=blue,citecolor=blue]{hyperref}
\usepackage{soul}
\allowdisplaybreaks

\pdfoutput=1 

\usepackage{graphicx}
\usepackage{hyperref}
\usepackage{aas_macros}
\usepackage{amsmath}
\usepackage{amssymb}
\usepackage{xcolor}
\usepackage{nccmath}
\usepackage{cleveref}

\newcommand{\pbh}{\text{PBH}}

\def\beq{\begin{equation}\begin{aligned}}
\def\eeq{\end{aligned}\end{equation}}

\begin{document}

\title{\boldmath Constraints on Primordial Black Holes from $N$-body simulations of the Eridanus II Stellar Cluster}

\author{Julia Monika Koulen,}
\email{jmkoulen@ucsc.edu}
\affiliation{Department of Physics, 1156 High St., University of California Santa Cruz, Santa Cruz, CA 95064, USA}
\affiliation{Santa Cruz Institute for Particle Physics, 1156 High St., Santa Cruz, CA 95064, USA}
\affiliation{Zentrum für Astronomie und Astrophysik, Technische Universität Berlin, Hardenbergstraße 36, D-10623 Berlin, Germany}

\author{Stefano Profumo,}
\email{profumo@ucsc.edu}
\affiliation{Department of Physics, 1156 High St., University of California Santa Cruz, Santa Cruz, CA 95064, USA}
\affiliation{Santa Cruz Institute for Particle Physics, 1156 High St., Santa Cruz, CA 95064, USA}

\author{Nolan Smyth}
\email{nwsmyth@ucsc.edu}
\affiliation{Department of Physics, 1156 High St., University of California Santa Cruz, Santa Cruz, CA 95064, USA}
\affiliation{Santa Cruz Institute for Particle Physics, 1156 High St., Santa Cruz, CA 95064, USA}

\begin{abstract}
\noindent The evolution of old, compact stellar structures provides strong constraints on macroscopic dark matter candidates such as primordial black holes. In view of recent observational data for the Eridanus II dwarf galaxy, we perform the first $N$-body simulations of its central stellar cluster to model dynamical heating by PBHs. We find evidence that such candidates must be lighter than about one solar mass if they constitute the totality of the dark matter. We additionally derive constraints on the fraction of the dark matter in macroscopic objects as a function of mass, by modeling the remainder of the dark matter as standard, fluid-like cold dark matter.
\end{abstract}

\maketitle
\newpage
\section{Introduction}
\label{sec:intro}

The nature of dark matter (DM), which modern cosmological models posit to constitute around $80\%$ of the non-relativistic matter in the Universe, remains mysterious and qualitatively unknown (for a constantly updated review see Ch.~27 of \cite{ParticleDataGroup:2022pth} and \cite{profumo_introduction_2017}). A large experimental effort pursuing weakly-interacting massive particles with mass around the weak scale has yielded only limits and no experimental evidence \cite{Arcadi_2018}. Concurrently, much attention has been placed on both more massive and much lighter DM candidates, such as primordial black holes \cite{Carr:2020gox,Carr:2020xqk} and light, bosonic, wave-like candidates \cite{Hui:2016ltb}.

Here, we focus on the most massive-possible DM candidates, a paradigmatic, but not unique, example of which is black holes of non-stellar origin, or primordial black holes (PBH) \cite{Carr:2020gox,Carr:2020xqk}. In the stellar-mass range, the fraction of cosmological DM that could be comprised of PBH is constrained by a number of considerations, which we review below. 

First, if stellar mass PBH are abundant, they are expected to form binaries and eventually merge, producing both a stochastic background of gravitational waves, as well as individual events detectable with gravitational-wave observatories such as LIGO/VIRGO/KAGRA \cite{VIRGO:2014yos,LIGOScientific:2014pky}. However, constraints based on gravitational wave production are subject to a number of uncertainties related to the rate of early and late binary formation, the clustering of PBH, and the possibility that some, or even most, of the LIGO/VIRGO/KAGRA events comprise PBH mergers themselves, as envisioned in \cite{Bird:2016dcv}.

A second class of constraints comes from direct microlensing searches. Here, new results from OGLE appear to place significant constraints \cite{Mroz:2024wag}. However, large uncertainties stem from computing the background event rate from dim stars or other astrophysical objects, as well as from modeling the Galactic DM halos and PBH clustering \cite{Mroz:2024wag}.

Lastly, stellar-mass PBH are slated to accrete particles in the interstellar medium (ISM), and in the process, accelerate them. This leads to measurable distortions in the cosmic microwave background spectrum \cite{Carr:1993aq}. Unfortunately, these constraints are again affected by uncertainties in the predicted accretion rate. This is due to a number of factors including whether an accretion disk forms, if accretion is spherically symmetric, and the relative velocity of the ISM and the PBH (see e.g.~\cite{DeLuca:2020fpg}).

Here, we focus on the gravitational effects of stellar-mass PBH on compact stellar clusters. We carry out detailed $N$-body simulations of stars, PBH, and fluid-like DM in excess of PBH, if the latter comprise only a fraction of the cosmological DM. In the past, similar studies utilized observations of the Eridanus II stellar cluster in the Eridanus II dwarf galaxy \cite{brandt_constraints_2016}, of the stellar cluster in the Leo T dwarf galaxy \cite{Lu:2020bmd}, and the Segue 1 galaxy \cite{Koushiappas:2017chw}. These previous analyses relied on semi-analytical methods. We focus on the Eridanus II (Eri II) system given that, compared to the data used in \cite{brandt_constraints_2016}, novel and critical observational data are now available on this system and on its central dense stellar cluster, in particular from Refs.~\cite{Crnojevi__2016, zoutendijk_muse-faint_2021}.

Eri II is a dwarf galaxy located in the vicinity of the Milky Way, just beyond our Galaxy's virial radius. It resides at a distance of $D = 366 \pm 17 ~ \textrm{kpc}$, with a half-light radius of $R_{\textrm{h}} =  277 \pm 14 ~ \textrm{pc}$ and houses a star cluster near its center with a half-light radius of  $r_{\textrm{h}} =  13 \pm 1 ~ \textrm{pc}$ \cite{2015ApJ...805..130K, DES:2015txk}. The dwarf galaxy was discovered relatively late in 2015 by the Dark Energy Survey (DES) \cite{DES:2015txk} and is classified as one of the Milky Way's satellite galaxies. The late discovery can likely be attributed to its relatively low luminosity; at an absolute magnitude of $M_V = -7.1$, Eri II is one of the least luminous galaxies known to contain a central star cluster \cite{Crnojevi__2016}. Ultra-faint dwarf galaxies, such as Eri II, possess limited stellar and gas content within a large DM halo, making them an excellent target for DM searches.
We focus on identifying the dynamical constraints for PBH by studying their gravitational effects on the central star cluster. Specifically, we study the PBH gravitational energy injection and the resulting heating of the central stellar cluster.

The remainder of this study is structured as follows:
In Sec.~\ref{sec:analytics} we review a semi-analytical approach to the problem; we then describe our numerical study setup in Sec.~\ref{sec:sim}, and present our findings in Sec.~\ref{sec:results}. The final Sec.~\ref{sec:conclusions} presents our discussion, conclusions, and outlook.

\section{Stellar Cluster Expansion from compact Dark Matter}
\label{sec:analytics}

Astrophysical black holes are formed by the gravitational collapse of massive stars or mass accumulations in galactic nuclei. However, black holes may also form in the early Universe as a result of the collapse of primordial density perturbations, cosmic strings, bubble collisions, etc.~\cite{Carr:2020gox, Carr:2020xqk}. 
Such black holes are dubbed \textit{primordial} black holes (PBH). For a review, see \cite{chapline_cosmological_1975, carr_primordial_2021}, although they may also have formed relatively recently (see e.g.~\cite{Picker:2023ybp}).

PBH may exist over a broad spectrum of masses, from the scale of supermassive black holes, with masses ranging from $10^6$ to $10^{10} ~ M_{\odot}$, to asteroid-masses around $10^{-16} ~ M_{\odot}$, a mass-scale at which the black hole lifetime from Hawking evaporation shortens to approximately the age of the Universe. The mass function of the PBH population depends on the formation mechanism but is often, for simplicity, taken to be monochromatic. Notice that for PBH formed by the collapse of symmetric density fluctuations, the mass distribution is expected to be described by a log-normal distribution, which approximates a delta function in the limit of very small width; for scale-invariant fluctuations, such as the collapse of cosmic strings, a power-law should result. Other mass functions are also possible (for details and references, the Reader should refer to Ref.~\cite{1975ApJ...201....1C, James_Turner_2020}).

Encounters between stellar systems and black holes primarily involve changes in the velocities or accelerations of stars. Tidal effects result from differential gravitational forces during close encounters between celestial bodies \cite{hills_possible_1975, bertotti_tidal_1990}. In encounters between stellar systems and black holes, these effects can lead to significant changes in the spatial distribution of stars, influencing the dynamics and structure of self-gravitating stellar systems. The extent of such tidal effects depends on various factors, such as the mass of the black hole, the distance between the black hole and the star cluster, and the arrangement of the stars within the star cluster.

As alluded to above, ultra-faint dwarf galaxies provide strong constraints on the abundances and masses of PBH. Dynamical heating leads to changes in the velocities of stars within the cluster, ultimately causing the expansion of the entire star cluster. The survival of the star cluster within the Eri II galaxy, a compact cluster inside a largely DM-dominated environment, therefore places limits on the abundance of PBH that would cause heating, as first pointed out in Ref.~\cite{brandt_constraints_2016}.

The process of gravitational heating by PBH of the star cluster can be modeled as a diffusion problem, with the sum of diffusion coefficients for a star's velocity describing the temporal evolution of its kinetic energy. For the purposes of analytic comparison, we take the DM particles to follow an isotropic Maxwellian velocity distribution and to have a locally uniform density \cite{profumo_introduction_2017}. Under these assumptions, the diffusion coefficient, which quantifies the degree of scattering in the velocity distribution of stars due to interactions with PBH, can be cast as
\begin{equation}
\label{eq:diffusion_coefficient}
    D[(\Delta v)^2] = \frac{4 \sqrt{2} \pi G^2\  f_\pbh \  \rho\ m_\pbh \  \ln{\Lambda}}{\sigma_\pbh}\bigg[\frac{\text{erf}(X)}{X}\bigg]\, ,
\end{equation}
where $m_\pbh$ and $\sigma_\pbh$ are the PBH mass and velocity dispersion, respectively, $\rho$ is the total DM density, $X \equiv \sigma_\star/\sqrt{2}\sigma_\pbh$\footnote{$\sigma_\star$ represents the typical stellar velocity, which we identify as the stellar velocity dispersion $\sigma_{\text{cluster}}$ of the cluster, such that $X$ quantifies the ratio between the stellar and PBH velocities.}, and $f_\pbh$ is the fraction of DM in PBH. $\textrm{ln} \Lambda$ is the Coulomb logarithm, quantifying the relative strength of long-range gravitational interactions versus short-range encounters among particles in a system \cite{binney_galactic_2008,brandt_constraints_2016}, and $G$ is Newton's constant.
The median Coulomb logarithm for point-masses interacting with extended substructures, as detailed in Ref.~\cite{penarrubiaOrbitalScatteringRandom2019}, is
\begin{equation}
    \label{eq:coulomb_ln}
    \text{ln} \Lambda \approx 8.2 \, .
\end{equation}



Under the assumption that the cluster resides in a DM core of constant density $\rho$, the potential energy of the cluster per unit mass can be expressed as
\begin{equation}
    \label{eq:pot_energy_unit_mass}
    \frac{U}{M} = \text{constant} - \alpha \frac{G M_\star}{r_{\text{h}}} + \beta G \rho r_{\text{h}}^2 \, ,
\end{equation}
where $M_\star$ represents the stellar mass of the cluster, and $\alpha$, $\beta$ are proportionality constants that depend on the mass distribution of DM.

A constant is present in the first term of Eq.~(\ref{eq:pot_energy_unit_mass}), serving as a reference point for the potential energy and essentially shifting the energy scale. The second term describes the gravitational energy of the star cluster, originating from the attraction among the stars within the cluster. The short-range effect of this term is given by the proportionality $\propto r_{\text{h}}^{-1}$, and $\alpha$ quantifies the strength of gravitational interactions within the cluster. The final term represents the potential energy induced by the DM surrounding the cluster.

A star cluster embedded in a DM core constitutes, on sufficiently long time-scales, a virial system. The virial theorem, in general, is given by
\begin{equation}
    \label{analystics:eq:virial_theorem}
    E_{\textrm{tot}} = \frac{1}{2} U \,.
\end{equation}
The temporal change of the potential energy per unit mass and the square of the mean velocity of the stars in the star cluster can be calculated as follows. For the potential energy per unit mass, after rearrangements and applying the chain rule, one obtains
\begin{align}
    \begin{split}
        \frac{dU(r_{\text{h}}(t))/M}{dt} &= \frac{dU(r_{\text{h}}(t))/M}{dr_{\text{h}}} \frac{dr_{\text{h}}}{dt} \\
        &= \left( \frac{\alpha G M_\star}{r_{\text{h}}^2} + 2 \beta G \rho r_{\text{h}} \right) \frac{dr_{\text{h}}}{dt} \, .
    \end{split}
\end{align}
As the diffusion coefficient describes the temporal evolution of a star's kinetic energy, it can be interpreted as the derivative w.r.t.~time of the squared mean velocity, with a diffusive component $D$. Further rearrangements yield an implicit equation for the temporal evolution of the star cluster's half-light radius, which expands due to dynamical heating caused by PBH as

\begin{equation}
    \label{eq:half_light_radius}
    \frac{d r_{\text{h}}}{dt} = \frac{4 \sqrt{2} \pi G\  f_{\text{PBH}}\ m_{\text{PBH}}\ \text{ln}\ \Lambda}{\sigma_\pbh} \bigg[\frac{\text{erf}(X)}{X}\bigg]\left( \frac{\alpha M_\star}{r_{\text{h}}^2 \rho} + 2 \beta  r_{\text{h}} \right)^{-1} \, .
\end{equation}


Recent HST photometry finds that Eri II's central cluster has an age of $13.5 \pm 0.3 ~\rm{Gyr}$ and a mass of $\sim 8.4 \times 10^{3} ~ M_{\odot}$, although the system may have been as much as $4$ times as massive at formation \cite{weiszReionizationeraGlobularCluster2023}. Observations yield a current half-light radius value of $r_{\text{h}} = 13 ~ \textrm{pc}$ for the star cluster \cite{Crnojevi__2016}. 


As an example, we take a cluster with a stellar mass of $M_\star = 8.4 \times 10^{3} ~ M_{\odot}$ and an initial half-light radius of $r_{\text{h}} = 1.0 ~ \text{pc}$. The DM core hosting the star cluster has a fraction of DM in the form of PBH with a mass of $m_{\text{PBH}} = 30 ~ M_{\odot}$ and a velocity dispersion of $\sigma_\pbh = 5~\mathrm{km\,s}^{-1}$.

Note that for e.g. a Plummer profile with $M_\star = 8.4 \times 10^3 ~ M_\odot$ and scale length $a\sim1.53 ~\text{pc}$, the central stellar velocity dispersion is approximately
$\sigma_{\text{cluster}}\sim\sqrt{\frac{GM_\star}{6a}}\sim 2 ~ \mathrm{km\,s}^{-1}$. Since $\sigma_{\mathrm{PBH}}=5 ~ \mathrm{km\,s}^{-1}$, we account for both scales by including the factor $\mathrm{erf}(X)/X$ in our heating rate, which has the net effect of slightly suppressing the energy injection and slows the expansion of the half-light radius relative to the results of \cite{brandt_constraints_2016}, which assumes $\mathrm{erf}(X)/X = 1$, i.e. the impulsive regime.

\begin{figure}[t]
\centering
\includegraphics[width=0.85\textwidth]{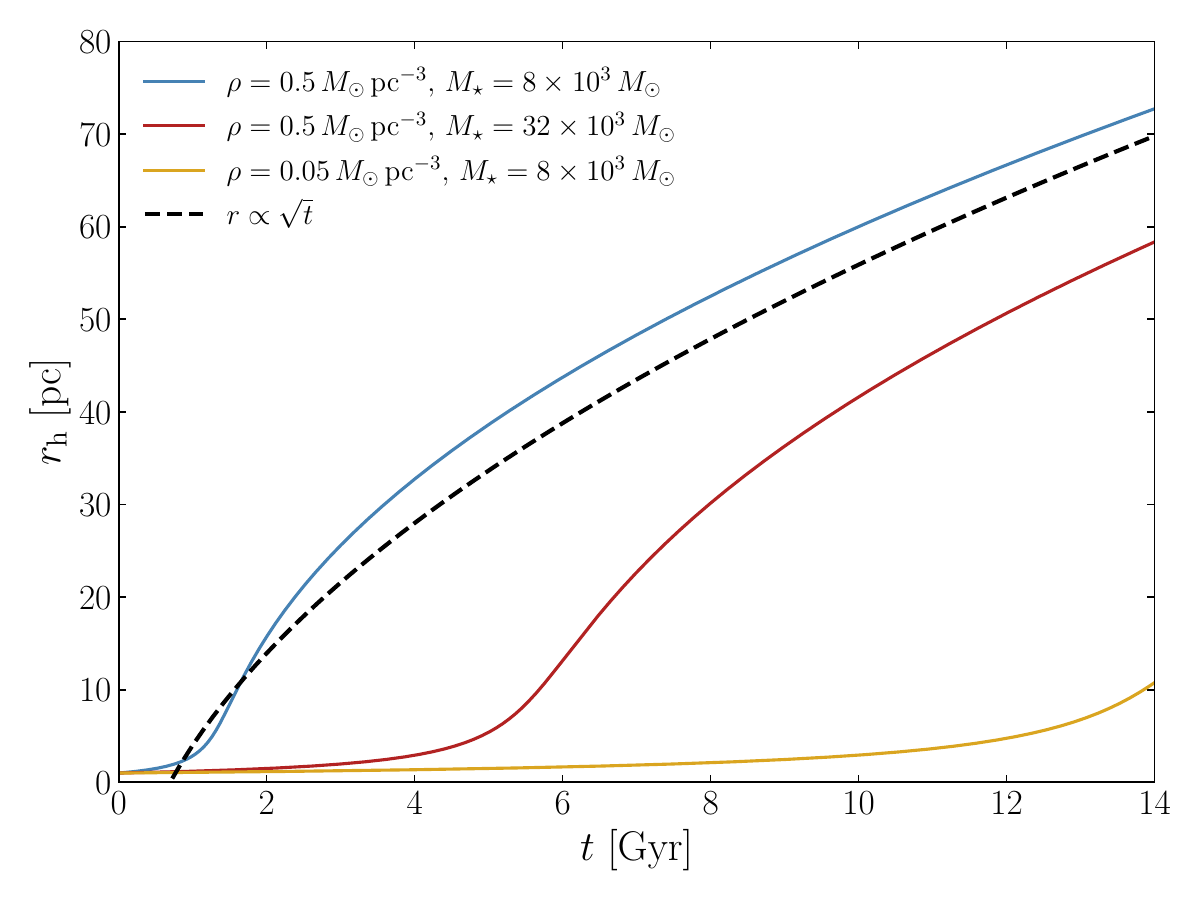}
\caption{\textbf{Evolution of the half-light radius over a time span of $14 ~ \text{Gyr}$.} The radius expands due to dynamical heating of the star cluster caused by PBH with masses $m_{\text{PBH}} = 30 ~ M_{\odot}$, PBH dispersion $\sigma_\pbh = 5 ~ \mathrm{km\,s}^{-1}$ and cluster velocity dispersion $\sigma_{\rm{cluster}} = 2 ~ \mathrm{km\,s}^{-1}$. The star cluster has an initial half-light radius of $r_{\textrm{h}} = 1 ~ \mathrm{pc}$. We show the evolution for different DM densities and star cluster masses, as indicated in the legend. Each curve is obtained by solving Eq.~(\ref{eq:half_light_radius}) for a PBH abundance of $f_{\text{PBH}} = 1.0$.}
\label{fig:radius_evolution_tsit}
\end{figure}

Figure \ref{fig:radius_evolution_tsit} shows the evolution of the half-light radius over $14 ~ \text{Gyr}$ for different values of the DM density and stellar cluster mass. We use the parameter values $\alpha = 0.4$ and $\beta = 10.0$ following \cite{brandt_constraints_2016}.
The transfer of energy from PBH to the stars causes the cluster's expansion. Initially, the system undergoes a slow expansion until it eventually grows as $r_{\text{h}} \propto \sqrt{t}$. The half-light radius exhibits a more accelerated expansion at higher total DM densities. This behavior can be attributed to the increased presence of PBH in regions of higher DM densities, subsequently leading to a faster enhancement of the kinetic energies of the stars. Consequently, the half-light radius undergoes more rapid expansion in these regions compared to areas with a lower total DM density.

To derive constraints on PBH abundances for corresponding PBH masses, we compare the semi-analytical evolution of the stellar system in Eri II to the observed half-light radius and age of the star cluster. Based on the half-light radius, a characteristic lifetime can be defined as the duration required for the cluster to expand from the size of a typical galactic star cluster of $r_{\text{h,initial}} = 2 ~ \text{pc}$ to its current observed size of $r_{\text{h,final}} = 13 ~ \text{pc}$\footnote{A second characteristic lifetime is defined in \cite{brandt_constraints_2016} that is equivalent in the limit of a dark-matter dominated system. We exclusively focus on the characteristic lifetime stated here as it provides more conservative constraints in the intermediate regimes.}.

Requiring that this timescale or the duration over which dynamical heating occurs exceeds the actual age of the cluster allows us to constrain the PBH abundance by solving Eq.~(\ref{eq:half_light_radius}) for the half-light radius $r_{\text{h}}$. In the following, we assume a star cluster age of $13.5 ~ \text{Gyr}$ and a mass of $\sim 8 \times 10^{3} ~ M_{\odot}$ \cite{weiszReionizationeraGlobularCluster2023}, also allowing for the possibility that the mass is up to $4$ times greater at formation. When determining the constraints, values for the PBH abundance $f_{\text{PBH}}$ and mass $m_{\text{PBH}}$ are excluded for which the half-light radius value of the star cluster exceeds the value of $r_{\text{h,final}} = 13 ~ \text{pc}$ within $13.5 ~ \text{Gyr}$. 

\begin{figure}
\centering
\includegraphics[width=0.9\textwidth]{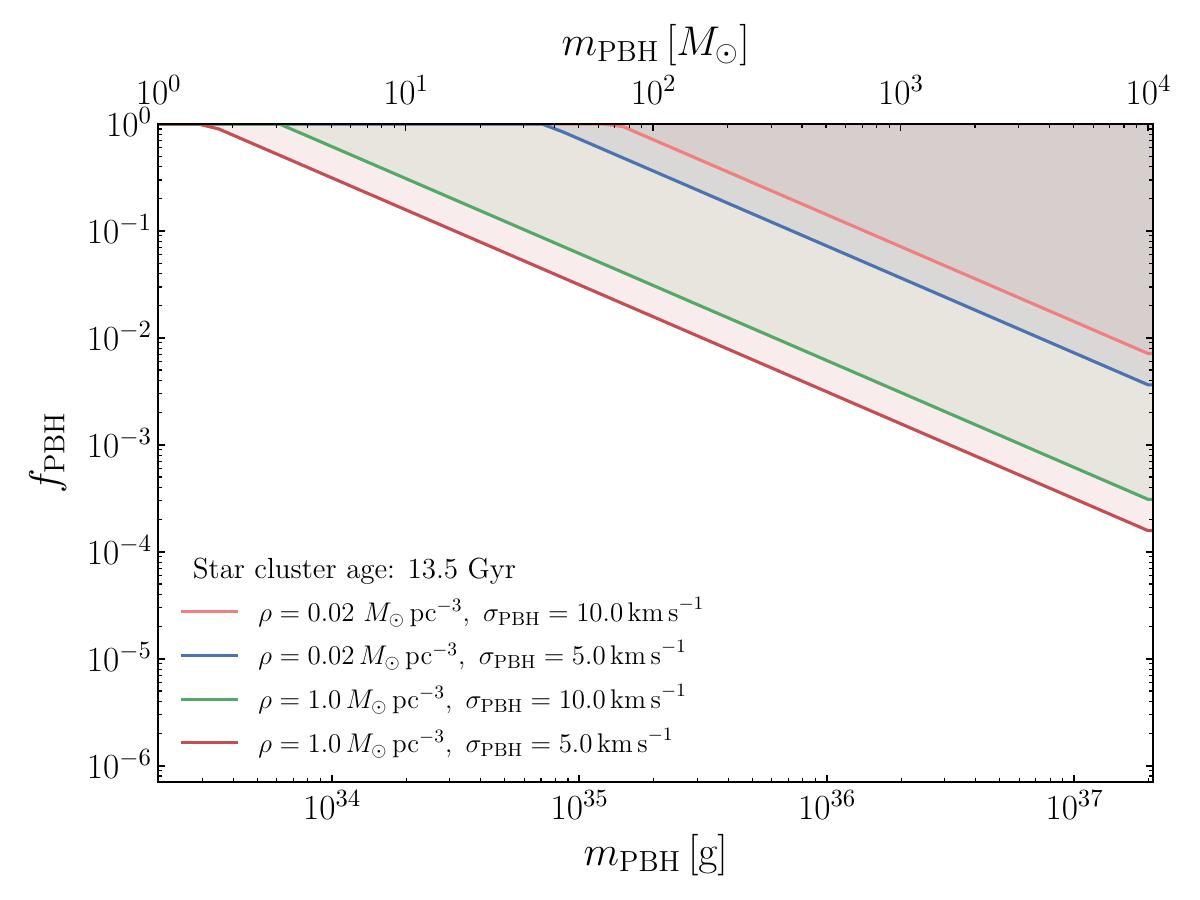}
\caption{\textbf{Semi-analytical constraints on PBH based on the survival of the star cluster in Eri II assuming a star cluster age of $13.5 ~ \text{Gyr}$.} Four different scenarios are represented with PBH densities $[0.02, 1.0] ~ M_{\odot}\,\text{pc}^{-3}$, PBH dispersion $[5.0, 10.0] ~ \mathrm{km\,s}^{-1}$ and stellar velocity of $2 ~ \mathrm{km\,s}^{-1}$. These scenarios involve PBH masses $m_{\text{PBH}}$ ranging from $1$ to $10^4 ~ M_{\odot}$. The constraints are derived by requiring that the time it takes the star cluster to grow from $r_{\text{h,initial}} = 2 ~ \text{pc}$ to its current observed size of $r_{\text{h,final}} = 13 ~ \text{pc}$ does not exceed $13.5 ~ \text{Gyr}$.}
\label{fig:constraints_Brandt_3}
\end{figure}

Figure \ref{fig:constraints_Brandt_3} shows the constraints on $f_{\text{PBH}}$ as a function of $m_{\text{PBH}}$ for a star cluster age of $13.5 ~ \text{Gyr}$. We consider different values of the three-dimensional DM velocity dispersion in the range of $[5, 10] ~ \mathrm{km\,s}^{-1}$ and densities of $[0.02, 1.0] ~ M_{\odot} \, \text{pc}^{-3}$. The values for the DM density and velocity dispersion are based on typical values for these quantities as found in ultra-faint dwarf galaxies \cite{simon_kinematics_2007,mcconnachie_observed_2012}.


For lower DM densities ($\rho = 0.02 ~ M_{\odot} \, \text{pc}^{-3}$), there are weaker constraints on PBH masses and abundances than for higher DM densities ($\rho = 1.0 ~ M_{\odot} \, \text{pc}^{-3}$). Slower-moving PBH $(\sigma_{\rm{PBH}} \approx 5.0 ~ \mathrm{km\,s}^{-1})$ interact with stars for a longer period of time and thus have a larger impact on the cluster. For higher PBH velocities ($\sim 10.0 ~ \mathrm{km\,s}^{-1}$), PBH must be more abundant to inject the same amount of energy, which is why the constraints are weaker for faster-moving PBH.


\section{Numerical setup and initial conditions}
\label{sec:sim}

In this section, we describe the suite of $N$-body simulations we performed using the package \texttt{GIZMO} \cite{hopkins_new_2015}. \texttt{GIZMO} is a code for cosmological $N$-body simulations of structure formation that uses smoothed particle hydrodynamics. \texttt{GIZMO} is derived from the \texttt{GADGET} code \cite{springel_cosmological_2005} and can be considered a part of the 3rd version of \texttt{GADGET} since it uses the parallelization scheme and Tree + PM gravity solver from \texttt{GADGET-3}, but additionally applies a Lagrangian meshless finite-mass (MFM) hydro solver.
The MFM method is an adaptive mesh-based method in which the system adapts to the matter density; it does not require a fixed regular mesh but rather kernels to compute the interactions between particles. This allows gas, stars, DM, and other astronomical objects to be simulated on a large scale. 

We run various simulation sets that are characterized by different PBH masses. Each simulation set contains subgroups that are distinguished by PBH abundance. 
Each subgroup consists of $50$ simulations, from which the median value is computed for each combination of PBH mass and abundance.
We run the simulations on the High-Performance Computing Cluster of the Math Institute of Technische Universität Berlin. The cluster comprises massively parallel processing computers, GPUs, and systems with particularly large main memory (CPUs: 16-Core Epyc 7302, 8-Core Epyc 7262, 16-Core Xeon 4216, HexaCore-Xeon E5-2630v2 and many more).

The \texttt{GIZMO} code uses parallelization via message passing interface, for which we use between $8$ and $64$ CPUs, depending on the particle number. The number of particles is also decisive for the duration of the simulation (as the runtime of these $N$-body simulations scales with $\mathcal{O}(N \log(N))$ \cite{appel_efficient_1985}). Thus, the simulations for a low number of particles take about $2$ hours and those for high numbers of particles up to $6$ days.

An $N$-body simulation models the motion of a large number of gravitationally interacting particles. Here, we consider background DM, PBH, and stars. We simulate the influence of PBH on stars within the Eridanus II galaxy to determine the extent and rate of expansion of the star cluster. As described above, we assume that the star cluster begins with an initial half-mass radius of $r_{\textrm{h,initial}} = 2 ~ \textrm{pc}$ and grows up to a maximum of $r_{\textrm{h,final}} = 13 ~ \textrm{pc}$ within $13.5 ~ \text{Gyr}$.

Note that we now shift from using the half-light to the stellar half-mass radius; this is done by exclusively considering the stellar mass in calculating $r_{\rm h}$ and assuming identical luminosity for each star. For consistency, we maintain the notation $r_{\textrm{h}}$ for the half-mass radius in this context, as it is assumed to be identical to the values discussed in Sec.~\ref{sec:analytics}.
We focus on the numerical output for the temporal evolution of the half-mass radius of the star cluster in the Eridanus II dwarf galaxy; the quantity of interest, $\frac{dr_\textrm{h}}{dt}$, will thus now be based exclusively on numerical simulation results and no longer on the solution of the differential equation as derived in the analytical approach in Sec.~\ref{sec:analytics}.

The simulation takes place in a non-cosmological framework, meaning it does not consider the Hubble expansion since it is irrelevant for a self-gravitating system on the scales of interest here. Additionally, a non-cosmological $N$-body simulation provides a controlled environment for isolating and understanding the specific interactions between PBH and stars.

To establish initial conditions for the numerical simulations, i.e., to generate the positions, velocities, and other relevant parameters for the system components (PBH, background DM particles, and stars), we use the package \texttt{DICE} \cite{2016ascl.soft07002P}. The initial conditions are computed with new random seeds for each simulation. We model the dwarf galaxy as spherically symmetric in the initial conditions. Although observations indicate that the star cluster of Eri II is located slightly off the center of the galaxy, our model assumes that it exists in the center (see Appendix \ref{sec:appendix} for details). We adopt a Sérsic profile \cite{1968adga.book.....S} for the density profile of the star cluster with a total mass of $M_\star = 8 \times 10^3 ~ M_{\odot}$ and with individual stellar masses of $m_\star = 1 ~ M_{\odot}$, resulting in $N_{\star} = 8 \times 10^3$.
The choice of these values for the stellar masses can be attributed to the description of the star cluster in \cite{weiszReionizationeraGlobularCluster2023, brandt_constraints_2016}. The half-mass radius of the star cluster is set to $r_{\textrm{h,initial}} = 2 ~ \textrm{pc}$ corresponding to the definition of the initial half-light radius of the characteristic lifetime. 

We use data from Zoutendijk et al.~\cite{zoutendijk_muse-faint_2021} to describe the DM halo of the Eri II galaxy. Zoutendijk et al.~employ stellar line-of-sight velocities gathered from observations of Eri II during the MUSE (Multi Unit Spectroscopic Explorer on the Very Large Telescope)-Faint survey to derive constraints on the DM profile. Their work enables the calculation of characteristic parameters for the shape of the density profile for various models of DM density profiles. For this study, we adopt their results and model the DM distribution using an NFW proifle with characteristic density $\rho_0$ and scale radius $r_{\textrm{s}}$ \cite{navarro_structure_1996}:
\begin{equation}
    \rho(r) = \frac{\rho_0}{\left(\frac{r}{r_\text{s}} \left( 1+\frac{r}{r_{\text{s}}} \right) \right)^2}.
\label{eq:NFW}
\end{equation}

As described in Sec.~\ref{sec:analytics}, halos such as that of Eri II can be assumed to have virialized within a virial radius $r_{\text{vir}}$. The virial mass of an NFW halo is \cite{navarro_structure_1996}
\begin{equation}
\label{eq:total_mass_NFW}
    M_{\textrm{vir,NFW}} = \int_{0}^{r_{\textrm{vir}}} 4 \pi \rho(r) dr = 4 \pi \rho_0 r_{\text{s}}^3 \bigg[ \textrm{ln}(1+c) - \frac{c}{1 + c} \bigg] \, ,
\end{equation}
where $c$ denotes the concentration parameter which is the ratio between the virial radius and the scale radius: $c = \frac{r_{\text{vir}}}{r_{\text{s}}}$.
The exact values for the individual galaxy parameters can be found in Tab.~\ref{tab:eriII_galaxy}.
By inserting the values in Tab.~\ref{tab:eriII_galaxy} into Eq.~(\ref{eq:total_mass_NFW}), we obtain the virial mass of the DM halo and thus the total mass of the system which is also listed in Tab.~\ref{tab:eriII_galaxy}.
\begin{table}
    \centering
    \begin{tabular}{|c|c|}
        \hline
        Quantity & Value \\
        \hline
        Characteristic density $\rho_0$ & $1.48 \times 10^8 ~ M_{\odot} \, \textrm{kpc}^{-3}$ \\
        Scale radius $r_{\textrm{s}}$ & $389 ~ \textrm{pc}$ \\
        Virial radius $r_{\textrm{vir}}$ & $9120 ~ \textrm{pc}$ \\
        Mass of an individual star $m_{\star}$ & $1 ~ M_{\odot}$ \\
        Mass of the stellar cluster $M_\star$ & $8 \times 10^3 ~ M_{\odot}$ \\
        Virial mass of the system $M_{\textrm{vir}}$ & $2.45 \times 10^8 ~ M_{\odot}$ \\
        \hline
    \end{tabular}
    \vspace{12pt}
    \caption{\textbf{Parameters used for the Eri II galaxy.} The values are taken or derived from \cite{zoutendijk_muse-faint_2021} and \cite{weiszReionizationeraGlobularCluster2023}.}
    \label{tab:eriII_galaxy}
\end{table}

For all simulations, the {\em total} mass of the DM halo is kept constant. Accordingly, to examine different PBH masses for varying PBH abundances, we adjust the number of PBH and of background DM particles. The value of $f_{\textrm{PBH}}$ is varied in the interval ${f_{\textrm{PBH}} = [10^{-4} - 1.0]}$, and the PBH mass within ${m_{\textrm{PBH}} = [10^3 - 10^4] ~ M_{\odot}}$ to compare the numerical results with those from the semi-analytical approach in Fig.~\ref{fig:constraints_Brandt_3}.



The mass of background DM and PBH particles is described by a monochromatic mass function, indicating that all PBH particles and background DM particles within a single simulation share the same mass, i.e., $m_{\textrm{PBH}} = m_{\textrm{BDM}}$. Thus, we consider three different particle types in each simulation: background DM particles, PBH, and the stars representing the star cluster at the halo's center.

We only investigate the gravitational interactions in the simulations, thus neglecting the largely unconstrained and poorly-understood effects of black hole accretion. When generating the initial conditions, the PBH and the background DM particles are both considered part of the DM halo. As such, the background DM and PBH particles not only have the same masses within a simulation set, but also the same velocity dispersion.

In $N$-body simulations, the softening length $\epsilon$ regulates numerical issues that arise at small scales. As the distance $r_{12}$ between two particles with masses $m_1$ and $m_2$ approaches zero, the gravitational force diverges as $\sim 1/r_{12}^2$, which can cause arbitrarily large accelerations and numerical instabilities. In physical systems, collisional effects and the finite extent of the masses would regulate this. Thus, $\epsilon$ is introduced to soften the gravitational force as $r_{12} \rightarrow 0 $ to prevent numerical artifacts \cite{rodionov_optimal_2005}.

\begin{equation}
    F_{\textrm{soft}} = \frac{G m_1 m_2}{r_{12}^2 + \epsilon^2}\, .
\end{equation}
For $r\gg\epsilon$ the expression $F_{\textrm{soft}}$ approaches the regular gravitational force $F$ and for $r\ll\epsilon$ the gravitational force approaches a maximal value for the two interacting masses. $\epsilon$ thus determines the distance below which the gravitational force between two particles is reduced, ensuring that it does not become infinitely large as the separation between particles approaches zero. The choice of an appropriate softening length is an important aspect of $N$-body simulations as it has a significant impact on the accuracy of the simulated astrophysical systems \cite{adamek_wimps_2019}. 

In our simulations, different particle types are distinguished by their characteristic softening lengths, reflecting their physical gravitational potential (e.g. $\epsilon_{\textrm{PBH}} \neq \epsilon_{\textrm{BDM}})$. To determine the softening lengths of the stars, $\epsilon_{\mathrm{stars}}$, and the background DM, $\epsilon_{\mathrm{BDM}}$, we first examine the case with no PBH ($f_{\mathrm{PBH}} = 0.0$). We select values that minimize the evolution of $r_\mathrm{h}$ over 13.5 Gyr considering the limiting case of the most massive BDM particles $m_{\mathrm{BDM}} = 10^4,M_{\odot}$.

PBH have effectively no macroscopic spatial extent, meaning the physical PBH size is very small compared to the distance between neighboring PBH. Thus we treat PBH as point particles with a small softening length $\epsilon_{\textrm{PBH}}$. We use Eq.~(\ref{eq:pbh_sep}) describing the typical separation between two neighboring PBH. It follows that the softening length of the PBH must be smaller than this length scale and the softening length of the background DM larger. In contrast, the background DM is expected to behave as a continuous medium that fills the background, reflecting its characteristic of a non-relativistic and collisionless fluid in the $\Lambda$CDM model. The softening length of the background DM must therefore be significantly larger than that of PBH to reflect the distinction between point-particle and fluid-like character, i.e., $\epsilon_{\textrm{BDM}} \gg \epsilon_{\textrm{PBH}}$. The softening lengths for the three particle types which fulfill the previously described criteria, and which are used in this study, are summarized in Tab.~\ref{tab:softening_lengths}.

\begin{table}
    \centering
    \begin{tabular}{|c|c|}
        \hline
        Softening length & Value \\
        \hline
        $\epsilon_{\textrm{BDM}}$ & $0.28 ~ \text{kpc}$\\
        $\epsilon_{\textrm{PBH}}$ & $1.9 \times 10^{-4}  ~ \text{kpc}$ \\
        $\epsilon_{\textrm{stars}}$ & $9.99 \times 10^{-5}  ~ \text{kpc}$ \\
        \hline
    \end{tabular}
    \vspace{12pt}
    \caption{\textbf{Values of the softening lengths for each particle type.}}
    \label{tab:softening_lengths}
\end{table}

Early simulations indicated that a single PBH with a low velocity near the cluster center could disproportionally dominate its early evolution. To mitigate this, we introduce an inner radial cut so that the evolution is due to the average affects of the total PBH population. This cut is based on the average separation between two neighboring PBH in the central region of the halo. We evaluate this at the characteristic density $\rho_0$, which is taken as a measure of the inner region of the DM halo. When calculating the inner radial density cut, we assume that each PBH particle is placed within individual volumes, denoted as $V_{\textrm{i}}$. Thus, the subvolume is defined as
\begin{equation}
    V_{\textrm{i}} = \frac{M_{\textrm{PBH}}}{\rho(r)} \, .
\end{equation}
Since the PBH abundances vary for each simulation, the characteristic density is multiplied by $f_{\textrm{PBH}}$ such that the density $\rho(r)$ at the center for different PBH concentrations is given by $\rho(r) = \rho_0 f_{\textrm{PBH}}$. 
Finally, we obtain the length scale associated with the volume of one PBH
\begin{equation}
\label{eq:pbh_sep}
    x = \Bigg( \frac{M_{\textrm{PBH}}}{\frac{4}{3}\pi \rho_0 f_{\textrm{PBH}}}\Bigg)^{\frac{1}{3}} \, .
\end{equation}

This length scale is adapted for the radius value at which the PBH start to appear for different $f_{\textrm{PBH}}$.
Introducing this cut prevents individual PBH from causing significant early expansion of the star cluster, as their isolated presence close to the cluster could result in a substantial energy transfer. 

\begin{figure}
\centering
\includegraphics[width=0.9\textwidth]{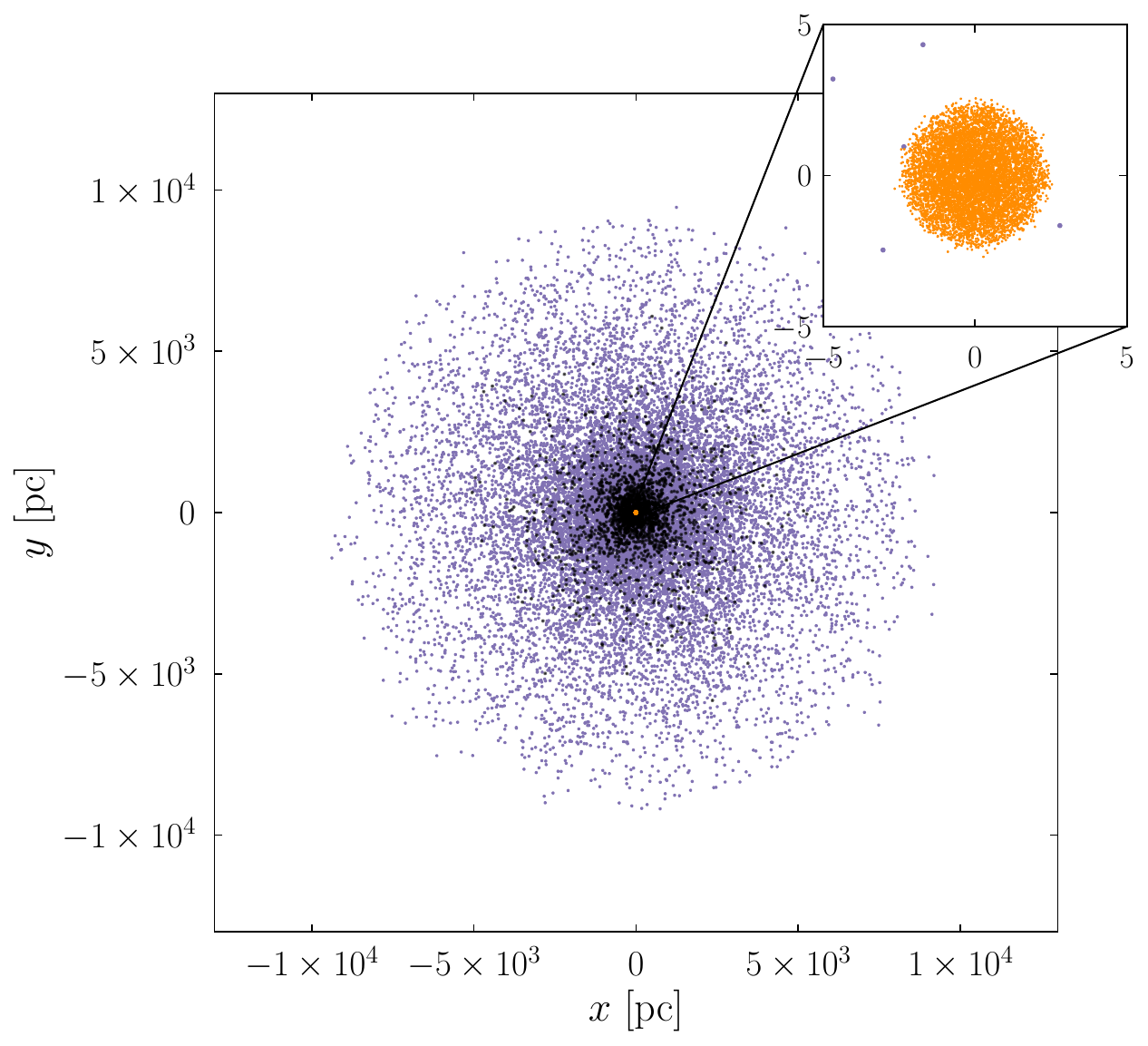}
\caption{\textbf{Snapshots of the initial condition state of the Eridanus II galaxy for $f_{\textrm{PBH}} = 0.1$ and $m_{\rm PBH} = 10^4 ~ M_{\odot}$.} Snapshot in the $xy$-plane. Background DM particles are indicated in purple, PBH in black, and stars in yellow. Upper right panel: Zoom-in snapshot of the simulation in the $xy$-plane. Both snapshots are at $t=0.0 ~ \text{Gyr}$.}
\label{fig:initial_conditions}
\end{figure}

Figure \ref{fig:initial_conditions} shows the initial conditions of a simulation, where the PBH and background DM masses are given by $m_{\textrm{PBH}} = 10^4 ~ M_{\odot}$, and the fraction of PBH is given by $f_{\textrm{PBH}} = 0.1$. Thus, this simulation includes $2,448$ PBH particles, $22,032$ background DM particles, and $8,000$ star particles. The purple particles represent the background DM, and the black particles represent PBH. The snapshot shows a view in the $xy$-plane, illustrating the entire system of the Eri II dwarf galaxy. The star cluster is represented by yellow particles in the galaxy's center. The same snapshot is shown in the upper right of Fig.~\ref{fig:initial_conditions} in a zoomed-in view. 

\section{Results}
\label{sec:results}

We first run sets of coarse grid simulations for logarithmically spaced PBH masses of $m_{\textrm{PBH}} = 10^3, \, 3\times 10^3, \, 10^4 ~ M_{\odot}$ with subgroups for the values in the interval $f_{\text{PBH}} = [10^{-4} - 1.0]$. This results in a total of $750$ simulations giving us an overview for the entire range of PBH abundance to be analyzed.

\begin{figure}
\centering
\includegraphics[width=0.9\textwidth]{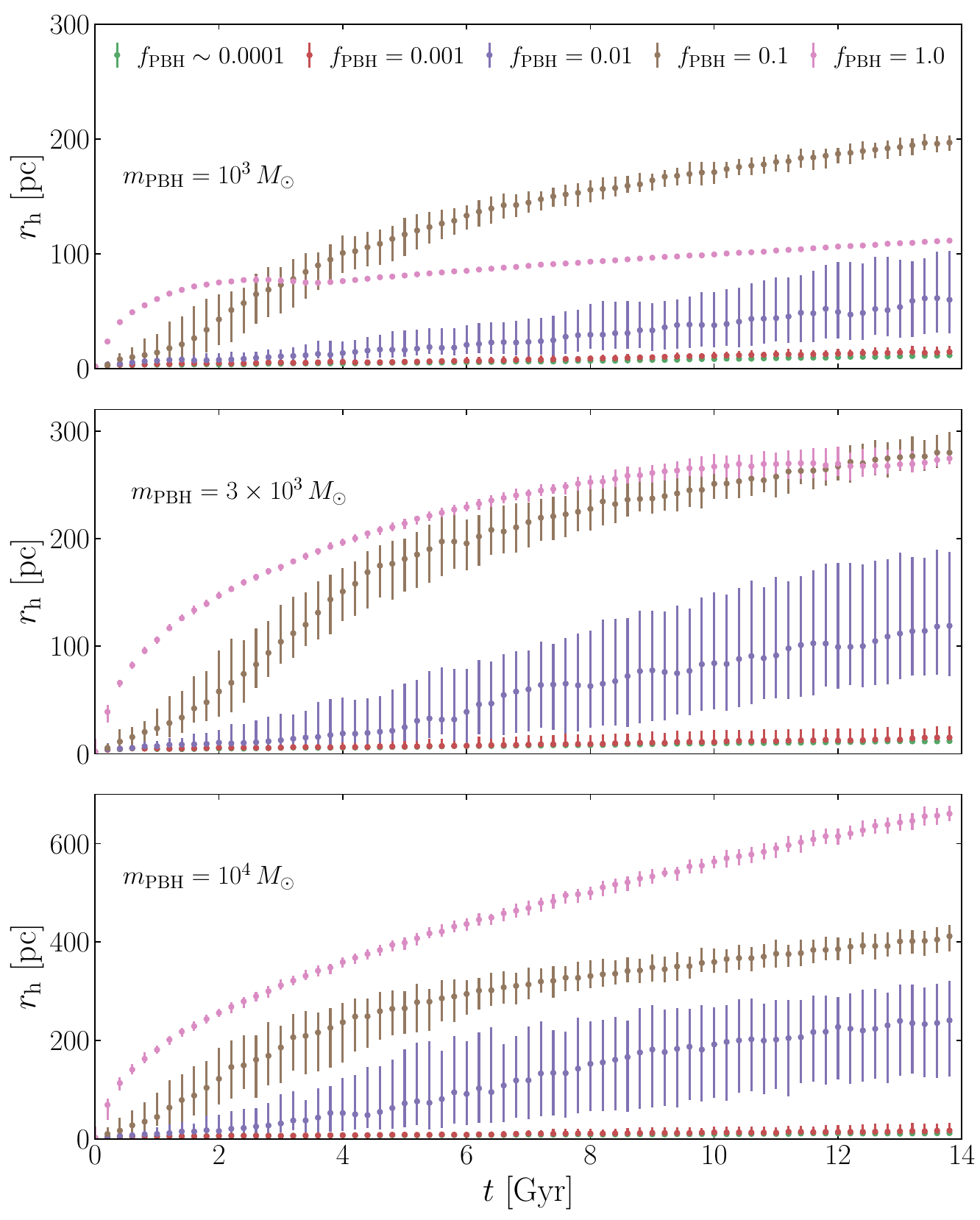}
\caption{\textbf{Evolution of the half-mass radius over $13.5 ~ \textrm{Gyr}$ for different PBH masses.} 
At each timestep and for each value of $f_\textrm{PBH}$, the curves show the median half-mass radius over 50 simulations, with error bars corresponding to the 16th–84th percentile interval.
}
\label{fig:radius_evolution_combined}
\end{figure}

The time evolution of the half-mass radius for the stellar component of the Eridanus II cluster is shown in Fig.~\ref{fig:radius_evolution_combined} for different PBH abundances. In the large majority of cases, the half-mass radius at any given time increases with $f_\text{PBH}$. This is because the larger number density of PBH results in an increased rate of dynamical heating interactions, driving the stars further from the center of the cluster.

The dynamical heating starts taking place at a rate that depends on $f_\text{PBH}$. This causes the stars to slowly expand from the center until the density becomes DM-dominated. At this point, the evolution of the half-mass radius scales as $r_{\rm h} \sim \sqrt{t}$ \cite{brandt_constraints_2016}. It is for this reason that the time evolution of the cluster is not strongly dependent on the PBH abundance for $f_\text{PBH} \sim 1$; if the number of PBH is great enough to rapidly heat the cluster within the first $\sim 2 ~\rm{Gyr}$, the evolution of the half-mass radius quickly reaches its asymptotic behavior. 

\begin{figure}
\centering
\includegraphics[width=0.9\textwidth]{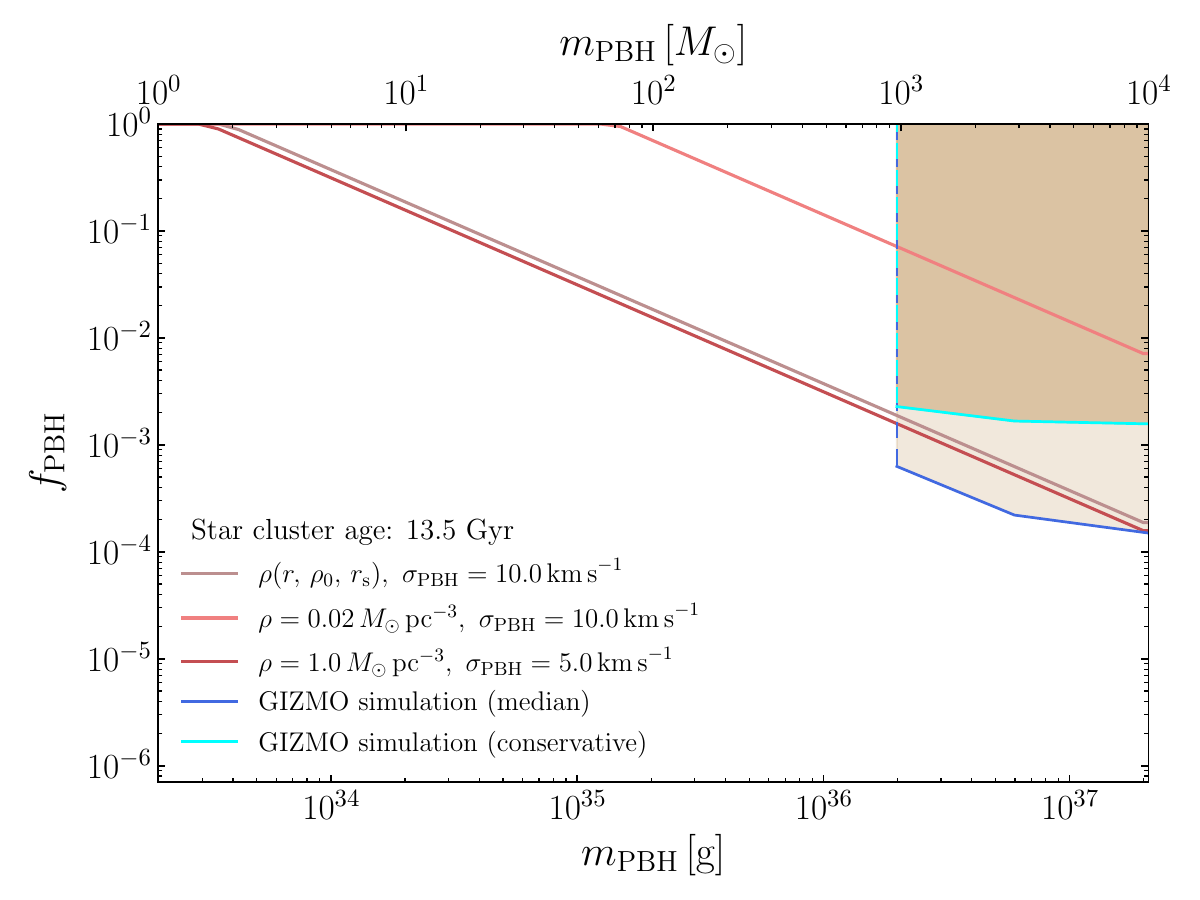}
\caption{\textbf{Constraints on $f_{\text{PBH}}$ as a function of $m_{\text{PBH}}$.} Limits are derived using the median $r_{\rm h}$ over $50$ simulations for each mass. Also shown is a conservative case using $r_{\rm h}$ one standard deviation below the median. The red and brown lines are the semi-analytically derived constraints.}
\label{fig:combined_constraints}
\end{figure}


In the following step, we run sets of refined grid simulations in which we examine the PBH abundance regions with smaller step sizes that are eligible for the constraints. 
The resulting limits on the abundance of PBH are shown as shaded regions in Fig.~\ref{fig:combined_constraints}. As in the analytic treatment in Sec.~\ref{sec:analytics}, these are obtained by determining the value of $f_\text{PBH}$ for which $r_{\rm h} \geq 13 ~\rm{pc}$ at $13.5 ~\text{Gyr}$. For larger abundances, the star cluster would have expanded beyond the size compatible with existing observations over its lifetime. We show both the limits derived using the median half-mass radius over 50 simulations, as well as those corresponding to the more conservative estimate using $r_{\rm h}$ one standard deviation below the mean value. In both cases, $r_{\rm h}$ is evaluated at $13.5 ~\text{Gyr}$ since it is monotonically increasing with time and thus only the last snapshot is needed to determine the corresponding limits on $f_\text{PBH}$. Note that $m_\text{PBH} = 10^3 ~ M_{\odot}$ is the lowest PBH mass for which a full numerical simulation was run due to computational efficiency and the large number density of particles associated with low masses ($n_{\text{PBH}} \propto f_{\text{PBH}} \frac{\rho_{\text{DM}}}{m_{\text{PBH}}}$). 

The red and brown lines show the semi-analytic bounds for both a constant density DM profile, as well as for an NFW density profile. As discussed in Sec \ref{sec:analytics}, a lower velocity dispersion yields more effective dynamical heating, resulting in more stringent constraints. This can be seen by comparing the two constant density curves. When comparing the NFW to the constant profile of identical dispersion $\sigma = 5 ~ \mathrm{km\,s}^{-1}$, the NFW profile results in a much stronger constraint. This is because of the enhanced DM density near the center of an NFW halo, which also increases the effects of dynamical heating. 

The constraints obtained from the $N$-body simulations are comparable to the semi-analytic results, with some key distinctions. While the semi-analytic curves exhibit similar behavior across the entire mass spectrum, the $N$-body simulations have a weaker mass dependence at the high-mass end.
Since the number of PBH decreases at higher PBH masses, very few PBH interact with the star cluster, resulting in fewer encounters between the star cluster and the PBH.
As a result, encounters between the PBH and the cluster are rare and stochastic, and the system enters a low-number, discrete-interaction regime compared to the semi-analytical results.
The relatively low number density of PBH at these large masses causes greater variation between runs, as evidenced by the difference between the conservative and median cases:
the flattening shown in the conservative curve should thus be ascribed to the noisy results of simulations with very few, massive PBH. At lower masses, the runs with larger number densities are more uniform and the conservative and median cases become comparable. Overall, the numerical results are more constraining than the semi-analytics across the entire mass range explored in this work.


\section{Discussion, Future Work, and Conclusions}
\label{sec:conclusions}

In this work, we examined the effects of PBH on the growth of stellar systems, placing dynamical limits on PBH as a DM candidate using cutting-edge $N$-body simulations. We first employed a semi-analytical approach to study dynamical heating of stellar systems due to  tidal effects from encounters with PBH. We focused on the specific example of the ultra-faint dwarf galaxy Eridanus II, which houses a central star cluster. The interaction with PBH dynamically heats the stars, leading to an expansion of the cluster's half-light or half-mass radius \(r_{\text{h}}\). We demonstrate how \(r_{\text{h}}\) evolves as a function of the mass and abundance of the PBH to impose dynamical constraints. As expected, we determined that a low PBH velocity dispersion imposes the strictest constraint for a given $f_\text{PBH}$. 

We then modeled the Eridanus II system within a non-cosmological $N$-body simulation comprised of stars, PBH, and background DM particles. We differentiated between PBH and background DM particles by treating PBH as point-like DM particles and background DM particles as fluid-like, with these distinctions manifesting in their gravitational potential ranges. 

Following the semi-analytic approach, we determined the star cluster's half-mass radius after $13.5~\text{Gyr}$. For each choice of $f_\text{PBH}, m_\text{PBH}$, we computed the median $r_{\mathrm{h}}$ over 50 simulations with different initial conditions. The requirement that $r_{\rm h} < r_\text{crit}$ results in constraints on $f_\text{PBH}$ for a given $m_\text{PBH}$.
Due to computational efficiency and the large number density of particles associated with low PBH masses, the mass range studied is limited to  $[10^3 - 10^4] ~ M_{\odot}$.

There are also some additional considerations and potential areas of future exploration that should be noted. When one particle type is much more massive than another, spurious heating effects can occur, leading to a systematic increase in the velocities of the less massive particle type \cite{wilkinson_impact_2023}. Here, however, we were concerned with cases where a physical PBH may be much more massive than the individual stars with which it interacts, requiring simulation of this regime. There may be ways to evaluate and mitigate the extent of such heating effects, which we leave to future investigation. 

We note that our fiducial setup assumes the stellar cluster lies at rest at the center of the Eridanus II dark-matter halo. This configuration is intentionally conservative because heating by primordial black holes depends on the local dark-matter density and velocity dispersion; both decline away from the center. At higher eccentricities, the cluster would briefly experience stronger tidal fields at pericenter, which could enhance energy injection and also accelerate mass loss. However, such configurations are short-lived. Therefore, our assumption of a static, central orbit yields conservative constraints.

Additionally, here we used the tree-based solver implemented in \texttt{GIZMO} to reduce the computational cost from $N^2$ to $N \log{N}$. This code was designed for collisionless dynamics, where numerical errors in the force calculation are assumed to be random. For the dense environments of stellar clusters, the typical numerical errors may be correlated due to the frequent, close-range interactions of stars and PBH. A direct summation calculation with a higher-order integration scheme could be used to provide higher numerical accuracy as done in \cite{stegmann_improved_2020} using the code \texttt{PHASEFLOW} \cite{vasiliev_new_2017}. At present, the computationally efficient Tree + PM gravity solver provides the first numerical constraints on PBH from Eri II as a baseline. Our results are broadly stronger than their semi-analytical counterparts, and significantly more reliable; compared with the results of Ref.~\cite{Lu:2020bmd},  based on the Leo T dwarf, our results here are stronger at low masses, and they are comparable to or more constraining than those from the Segue 1 galaxy \cite{Koushiappas:2017chw}.

\acknowledgments
This material is based upon work supported in part
by the National Science Foundation Graduate Research
Fellowship under Grant No. DGE-1842400 (NS). This work is partly supported by the U.S.\ Department of Energy grant number de-sc0010107 (SP). 

\appendix
\section{Introduction of a subhalo and a shift for the star cluster}
\label{sec:appendix}
\begin{figure}
\centering
\includegraphics[width=0.8\textwidth]{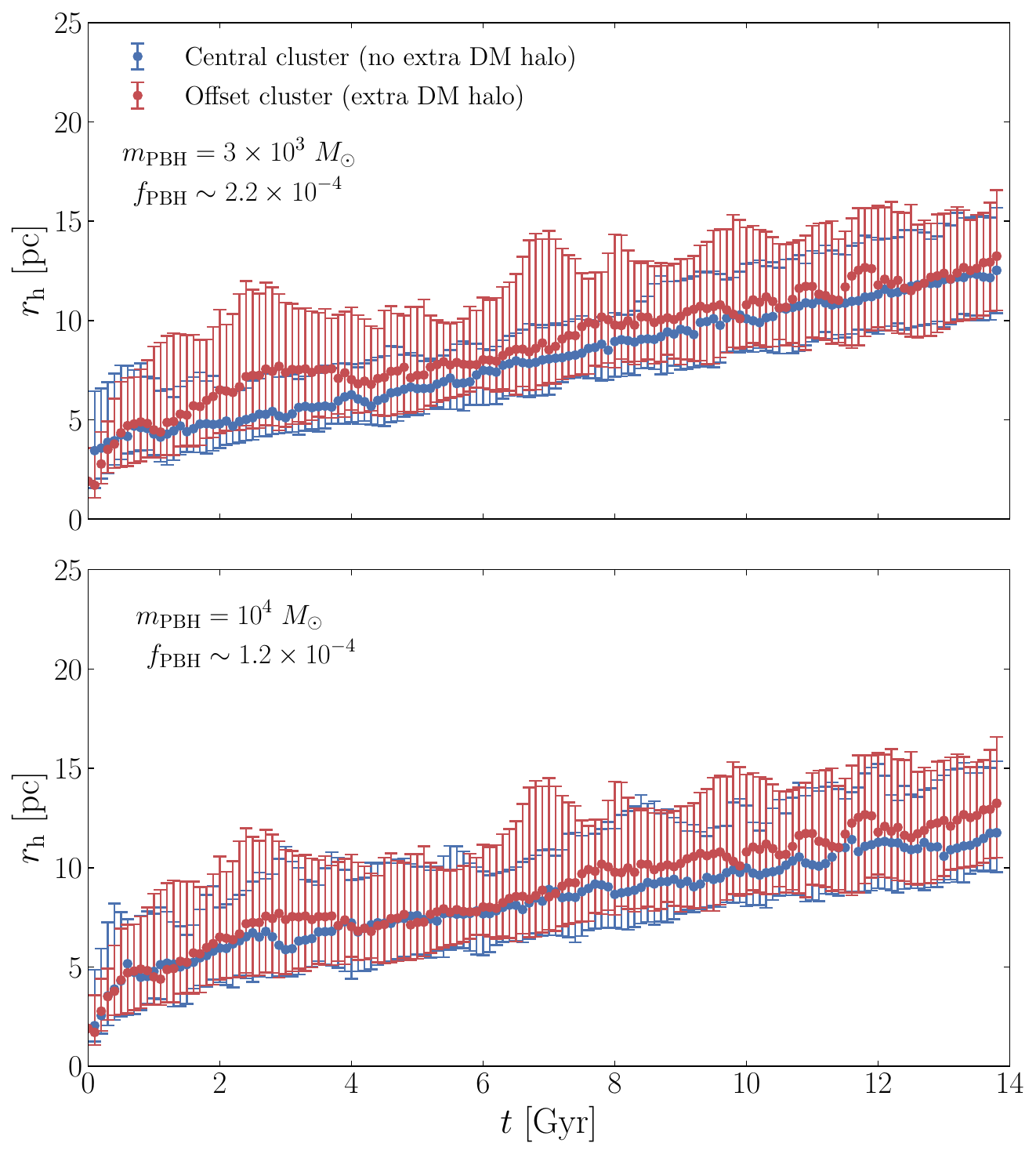}
\caption{\textbf{Evolution of $r_{\text{h}}$ over $13.5 ~\text{Gyr}$ for models without (top) and with (bottom) an additional dark matter halo and an initially offset star cluster.} 
At each timestep, the curves show the median half-mass radius over 50 simulations for both the central cluster without an additional DM halo (red) and the offset cluster with an additional DM halo (blue), with error bars corresponding to the 16th–84th percentile interval.}
\label{fig:comparison_regular_special_case}
\end{figure}

In \cite{penarrubia_capture_2024}, Eri II is identified as a galaxy with unusually low luminosity, metallicity, and size, whose properties could be explained by the presence of an additional DM halo surrounding its central star cluster.
Furthermore, it is shown in \cite{simon_eridanus_2021}, that the cluster has an offset from the galaxy center by a projected distance of $\sim 23~\text{pc}$, corresponding to a distance of approx.~$40~\text{pc}$.
We investigate the combination of these two scenarios by running 50 additional simulations 
for each of the points that define the constraint for PBH masses of   and $m_{\text{PBH}} = 10^4~M_{\odot}$ (see Fig.~\ref{fig:combined_constraints}), in order to examine whether the presence of an additional DM halo and the cluster’s offset have any influence on the results previously obtained.
These simulations include an additional DM halo of $\sim 10^6~M_{\odot}$ around the star cluster, which has a displacement of $40~\text{pc}$.
Following \cite{penarrubia_capture_2024}, we model the additional DM halo as a compact substructure with a scale radius of $\sim 3.78 ~ \text{pc}$ and a Hernquist density profile \cite{hernquist_analytical_1990}.
The effect of the substructure and the star cluster offset is analyzed based on the evolution of the half-mass radius of the star cluster over its lifetime. The results are shown in Fig.~\ref{fig:comparison_regular_special_case}, where the red curve represents the half-mass radius evolution for the case in which the additional DM halo and offset are present, while the blue curve corresponds to the evolution without these two properties and thus to the constraints shown in Fig.~\ref{fig:combined_constraints}.
In all cases, the evolution of the cluster has not changed significantly, with the distribution of half-value radii matching that of the non-shifted and halo-free simulation suites with the same PBH mass and abundance.
Based on this, we conclude that an additional DM halo and an offset of the star cluster would not have a noticeable effect on or change the derived constraints in this work.

\bibliographystyle{apsrev4-1}
\bibliography{references}
\end{document}